\documentclass[11pt,fleqn]{llncsToc}
\usepackage{latexsym,ifthen,amssymb,array,stmaryrd}
\usepackage[latin1]{inputenc}
\usepackage{a4wide}
\newcommand{\And}{\mathrel{\wedge}} 
\newcommand{\Or}{\mathrel{\vee}} 
\newcommand{\False}{\mti{false}} 
 
\newcommand{\Imp}{\Rightarrow} 
 
\newcommand{\Totfunc}{\rightarrow}

\newcommand{\Com}[2]{\mbox{\Esp{-1em}$#1$\Esp{1em}%
    \ifthenelse{\equal{#2}{ }}{}{\makebox[.9\textwidth]{\mbox{}\hfill$\{$ #2 $\}$}}}}
\newenvironment{Preuve}{\[\begin{array}{>{$}p{.9\textwidth}<{$}}}{\end{array}\]} 

\newcommand{\mti}[1]{\mbox{{\it #1}}}

\newcommand{\open}{\;\widehat{\,}\;}

\newcommand{\cpl}{\overline}
\newcommand{\ets}{\varnothing}

\newenvironment{PreuveL}{\[\begin{array}{>{\mbox{\Esp{-2ex}}\hfill}%
p{2.5ex}<{.\Esp{1ex}}>{\Esp{-1ex}$}p{.66\textwidth}<{$}>{;\ }p{.25\textwidth}}}{
\end{array}\]}

\newcommand{\Nat}{\mathbb{N}}
\newcommand{\Pow}{\mathbb{P}}

\newcommand{\dom}{\mbox{\textsf{dom}}}

\newcommand{\fix}{\mbox{\textsf{fix}}}
\newcommand{\FIX}{\mbox{\textsf{FIX}}}
\newcommand{\pre}{\mbox{\textsf{pre}}}

\newcommand{\str}{\mbox{\textsf{str}}}

\newcommand{\grs}{\mbox{\textsf{grd}}}
\newcommand{\grd}{\mathit{grd}}

\newcommand{\Rarrow}{\Longrightarrow}

\newcommand{\bop}{}
\newcommand{\eop}{\mbox{}\hfill $\Box$}

\newcommand{\UNLESS}{\mathrel{\BF{unless}}}

\newcommand{\Skip}{\mti{skip}}
\newcommand{\Sel}{\mathrel{[\mbox{\hspace{-.25ex}}]}} 
\newcommand{\sco}{\ensuremath{\mathrel{\mbox{\bf ;}}}}
\newcommand{\Def}{\mathrel{\widehat{=}}}
\newcommand{\m}[1]{\ensuremath{\mathit{#1}}}
\newcommand{\Esp}[1]{\mbox{\hspace*{#1}}}
\renewenvironment{PreuveL}[1]{\[\begin{array}{>{\mbox{\Esp{-2ex}}\hfill}%
      p{2.5ex}<{.\Esp{1ex}}>{\Esp{-1ex}$}p{.66\textwidth}<{$}>%
      {\ifthenelse{\equal{#1}{N}}{ }{;\ }}%
      p{.25\textwidth}}}{\end{array}\]}
\newcommand{\dtl}{\mathrel{\triangledown}}
\newcommand{\B}{{\sf B}}
\newcommand{\Li}{{\cal L}}
\newcommand{\Cf}{{\cal F}}
\renewcommand{\bop}{\textbf{Proof}\hfill}
\renewcommand{\eop}{\mbox{}\hfill $\square$}
\newcommand{\PP}{{\cal P}}
\newcommand{\QQ}{{\cal Q}}

\titlerunning{Proof Obligations for Liveness Properties under Weak Fairness}
\begin{document}
\thispagestyle{empty}
\parbox{.8\textwidth}{ %
  {\Huge\bf IMAG}\\\\
  {\large\bf Institut d'Informatique et de \\[.5ex]
    Math\'{e}matiques Appliqu\'{e}es \\[.5ex]
    de Grenoble} 
}\hfill
\vspace{3cm}
\parbox{\textwidth}{%
  \begin{center}
    {\Huge\bf LSR}\\[1ex]
    {\LARGE\bf Laboratoire Logiciels, Syst\`{e}mes, R\'{e}seaux}
  \end{center}
}
\vfill
\begin{minipage}{\textwidth}
\title{%
  {\LARGE\bf RAPPORT DE RECHERCHE}\\[3ex]
  Proof Obligations for Specification and Refinement of
  Liveness Properties under Weak Fairness}
\author{H\'{e}ctor Ru\'{\i}z Barradas\inst{1,2} \and Didier Bert\inst{2}}
 \institute{Universidad Aut\'{o}noma Metropolitana Azcapotzalco, M\'{e}xico
   D. F.,  M\'{e}xico\\
 \email{hrb@correo.azc.uam.mx, Hector.Ruiz@imag.fr}
 \and
 Laboratoire Logiciels, Syst\`{e}mes, R\'{e}seaux - 
 LSR-IMAG - Grenoble, France\\
 \email{Didier.Bert@imag.fr}
 }
\maketitle
\end{minipage}
\vfill
\parbox{\textwidth}{%
RR 1071-I LSR 20 \hfill F\'{e}vrier 2005 \\[3ex]
\begin{center}
  {B.P. 72 - 38402 SAINT MARTIN D'HERES CEDEX - France}\\[2ex]
  {Centre National de la Recherche Scientifique}\\[1ex]
  {Institut National Polytechnique de Grenoble}\\[1ex]
  {Universit\'e Joseph Fourier Grenoble I}
\end{center}}
\newpage
\mbox{}
\newpage
\thispagestyle{empty}
\newpage
\thispagestyle{empty}
\begin{center}
\Large\bf Proof Obligations for Specification and Refinement of\\
  Liveness Properties under Weak Fairness
\end{center}
\vfill
\section*{Abstract}
In this report we present a formal model of fair iteration of events
in a \B\ event system. The model is used to justify proof obligations
for basic liveness properties and preservation under refinement of
general liveness properties. The model of fair iteration of events
uses the dovetail operator, an operator proposed in \cite{BrNeAFCD} to
model fair choice. The proofs are mainly founded in fixpoint
calculations of fair iteration of events and weakest precondition
calculus.

\subsection*{Keywords}
Liveness properties, Event systems, \B\ method, 
Unity logic, Refinement, Fairness, Weak Fairness.

\section*{R\'esum\'e}
Dans ce rapport nous pr\'esentons un mod\`ele formel d'it\'eration
\'equitable d'\'ev\'enements dans un syst\`eme \B\ \'ev\'enementiel.
Le mod\`ele est utilis\'e pour justifier des obligations de preuve des
propri\'et\'es de vivacit\'e de base et de la pr\'eservation dans les
raffinements des propri\'et\'es de vivacit\'e g\'en\'erales. Le
mod\`ele d'it\'eration \'equitable d'\'ev\'enements utilise
l'op\'erateur \emph{dovetail\/}, un op\'erateur propos\'e dans
\cite{BrNeAFCD} pour mod\'eliser une s\'election \'equitable. Les
preuves sont fond\'ees principalement sur des calculs de point fixe de
l'it\'eration \'equitable d'\'ev\'enements et le calcul des plus
faibles pr\'econditions.

\subsection*{Mots-cl\'es}
Propri\'et\'es de vivacit\'e, syst\`eme d'\'ev\'enements, m\'ethode \B, 
logique Unity, rafinement, \'equit\'e, \'equit\'e faible.
\vfill
\mbox{}
\newpage
\mbox{}
\newpage
\setcounter{tocdepth}{2}
\mbox{}
\vfill
\begin{minipage}{\textwidth}
\tableofcontents
\end{minipage}
\vfill
\mbox{}
\newpage
\mbox{}
\newpage
\pagestyle{plain}
\setcounter{page}{1}
\section{Introduction}

In \cite{RuBePDHE} we proposed the specification and proof of liveness
properties under a weak fairness assumption in \B\ events systems
\cite{AbrMusIDCB}. The syntax and semantic of liveness properties that
we adopted are similar to the ones used in {\sc unity} \cite{ChMiPPD}. 

Liveness properties are divided in two classes: basic liveness
properties and general liveness properties. Basic properties are
specified by the \emph{ensures\/} relation $\gg_w$. General liveness
properties are specified by the \emph{leads to\/} relation $\leadsto$.
$\gg_w$ and $\leadsto$ are relations between predicates on the system
state.

We proposed two proof obligations for basic liveness properties
founded on weakest precondition calculus. The proof of general
liveness properties is made by applying inference rules of the {\sc
  unity} logic.

Following the \B\ method, an abstract model can be refined in a more
concrete one. To preserve through refinement liveness properties
specified in abstract models, we proposed two other proof obligations.
One proof obligation is discharged by applying weakest precondition
calculus, an the other one need to identify basic liveness properties
in the refinement and to apply the {\sc unity} logic.

The goal of this report is to justify the proof obligations concerning
proofs of basic liveness properties and preservation of general
liveness properties under refinement, by a reasoning on the set
theoretic formulation of event systems. Our approach was inspired by
\cite{AbrMusIDCB}, where proof obligations concerning modalities are
justified by fixpoints of iteration of events, instead of a reasoning
over the set of traces in a system, as it is done in \cite{BaQiRFAS}.
However, our approach uses a model including a fair choice operator,
which allows us to model our weak fairness assumption over the
iteration of events. 

This report is structured as follows. In section \ref{dovetail} we
present the main definitions used in this work. In particular we
define the liberal set transformer for events in a \B\ system and we
present the dovetail operator which is used to model our fairness
assumption. In section \ref{BLP} we introduce the proof obligations
for basic liveness properties and we prove that they are sufficient
conditions to guarantee that fair iteration of events in the system,
terminates in a state satisfying the postcondition established by the
basic liveness property. In section \ref{GLP} we present how to
specify and prove general liveness properties. Moreover, we give two
proof obligations to guarantee preservation of general liveness
properties under refinement, and we demonstrate they are sufficient
conditions to ensure that fair iteration of refined events terminates
into a state satisfying the predicate established by the general
liveness property. In section \ref{CONC} we give the conclusions of
this report and some comments about the future work.

\section{The dovetail operator}\label{dovetail}

In \cite{BrNeAFCD} the dovetail operator, a fair nondeterministic choice
operator, is introduced. In this section we give the definition of this
operator by its weakest liberal transformer. In the first part of this
section we define the weakest liberal set transformer of events in a \B\
event system. In the second part we give the formal definition of the
dovetail operator by definition of its weakest liberal set transformer
and its termination set. 

\subsection{The Liberal Set Transformer}\label{Lib_set}

In \cite{AbrTBB}, each generalized substitution $S$ has associated a set
transformer $\str(S)$ of type $\Pow(u) \Totfunc \Pow(u)$, where $u$ is the
state space of a machine or refinement. For any $r$ in $\Pow(u)$,
$\str(S)(r)$ denotes the largest subset of states where the execution of
$S$ must begin in order for the substitution $S$ to terminate in a state
belonging to $r$. In \cite{AbrMusIDCB}, the events of a \B\ system are
formalized by conjunctive set transformers, but instead of identifying the
set transformer associated with an event $F$ by $\str(F)$, it is denoted by
its name $F$. In this way $F(r)$ denotes the set $\str(F)(r)$, where $F$ is
an event of a \B\ system and $r$ a subset of the state space $u$. In what
follows, we use this notation.

In order to deal with the notion of the weakest liberal precondition of an
event $F$ we define \emph{the liberal set transformer\/} of an event $F$ as
$\Li(F)$.
\begin{definition}\label{Li}
  \mbox{The Liberal Set Transformer} \[\Li(F) = \lambda r \cdot
  (r\in\Pow(u)\,|\, \{x\,|\,x\in u \And \m{wlp}(F,x\in r)\}) \]
\end{definition}

The set $\Li(F)(r)$ denotes the largest subset of states where the
execution of event $F$ must begin in order for $F$ to terminate in a state
belonging to $r$ or loop. The liberal set transformer of the events in a
\B\ system are defined as follows:
\[
\hspace{-\mathindent}
\parbox{\textwidth}{
  \begin{tabular}{ll}
  \parbox{.6\textwidth}{
    \[\begin{array}{l}
      \Li(\Skip)(r) = r \\
      \Li(F\Sel G)(r) =\Li(F)(r)\cap \Li(G)(r)\\
      \Li(p\;|\;F)(r) = (p \cup \{x\;|\;x\in u \And u\subseteq
      r\})\cap \Li(F)(r)
    \end{array}\]
  }
  &
    \parbox{.4\textwidth}{
    \[\begin{array}{l}
      \Li(F\sco G)(r) = \Li(F)(\Li(G)(r)) \\
      \Li(p \Rarrow F)(r) = \cpl{p}\cup \Li(F)(r) 
    \end{array}\]
    }
\end{tabular}}
\]
where $r$ and $p$ are subsets of $u$ and $\cpl{p}$ is $u-p$. We note,
that set $\{x\;|\;x\in u \And u\subseteq r\}$ in the liberal set
transformer of the preconditioned event may have only two values:
$\ets$ for $r\not = u$ or $u$ for $r=u$.  In the guarded command $p
\Rarrow F$ and the preconditioned event $p\;|\;F$, we follow the
notation introduced in \cite{AbrMusIDCB} where the guard or the
precondition of the commands is a set instead of a predicate.
Definitions of liberal set transformers presented here are the set
counterpart of definitions in \cite{SduIBRB}.

We remark that definitions of liberal set transformer of any set
transformer $S$, made up of set transformers $F$ or $G$, such that
$\Li(F)(u)=u$ and $\Li(G)(u)=u$, and operators $\Sel$, $|$, $\Rarrow$
and $\sco$, respect $\Li(S)(u)=u$ \cite{DijSch}.

The set transformers $F(r)$ and $\Li(F)(r)$ for event $F$ and postcondition
$r$ are related by the pairing condition:
\begin{equation}\label{paircond}
  F(r) = \Li(F)(r) \cap \pre(F)
\end{equation}
where $\pre(F)$, the termination set of $F$ is equal to $F(u)$.
From the pairing condition we conclude the implication:
\begin{equation}\label{LieqStr}
  F(u)=u \Imp F(r) = \Li(F)(r) \quad \mbox{for any $r$ in $\Pow(u)$}
\end{equation}
which indicates that the set transformer $F$ and $\Li(F)$ are the same
provided the event $F$ always terminates.
When $F(r)$ or $\Li(F)(r)$ are recursively defined:
\[F(r)  = {\cal F}(F(r)) \quad \mbox{or} \quad \Li(F)(r)  =  {\cal
  G}(\Li(F)(r)) \]
for monotonic functions ${\cal F}$ and ${\cal G}$, according to
\cite{HehDCO} we take $F(r)$ as the strongest solution of the equation $X =
{\cal F}(X)$ and $\Li(F)(r)$ as the weakest solution of the equation $X =
{\cal G}(X)$. As these solutions are fixpoints, we take $F(r)$ as the least
fixpoint of ${\cal F}$ ($\fix({\cal F})$) and $\Li(F)(r)$ as the greatest
fixpoint of ${\cal G}$ ($\FIX({\cal G})$).

\subsection{Definition of Dovetail Operator}

The dovetail operator is used to model the notion of fair scheduling of two
activities. Let $A$ and $B$ be these activities, then the operational
meaning of the construct $ A \dtl B$ denotes the execution of commands $A$
and $B$ fairly in parallel, on separate copies of the state, accepting as
an outcome any proper, nonlooping, outcome of either $A$ or $B$.  The fair
execution of $A$ and $B$ means that neither computation is permanently
neglected if favor of the other.

A motivating example of the use of the dovetail operator is given in
\cite{BrNeAFCD}. In that example the recursive definition:
$ X = (n:=0 \dtl (X\sco n:=n+1))$ 
which has as solution ``set $n$ to any natural number'', is contrasted
with the recursion $Y = (n:=0 \Sel (Y\sco n:=n+1))$ 
 which has as solution ``set $n$ to any natural number or loop''. The
possibility of loop in $X$ is excluded with the dovetail operator because
the fair choice of   statement $n:=0$ will certainly occur. In $Y$ the
execution of that statement is not ensured.

The semantic definition for dovetail operator in \cite{BrNeAFCD} is given
by definition of its weakest liberal precondition predicate transformer
($\m{wlp}$) and its termination predicate $\m{hlt}$. We give an equivalent
definition using the weakest liberal set transformer $\Li$ and its
termination set $\pre$:
\begin{definition}\label{Ldtl}
  \mbox{The Dovetail Operator}
  \[\begin{array}{l}
    \Li(F\dtl G)(r) = \Li(F)(r)\cap\Li(G)(r) \\
    \pre(F\dtl G)  =  (F(u)\cup G(u))\cap (\cpl{F(\ets)} \cup G(u))
    \cap (\cpl{G(\ets)}\cup F(u)) \\
    \pre(F\dtl G) = (F(u)\cap G(u)) \cup (\cpl{F(\ets)}\cap F(u)) \cup
    (\cpl{G(\ets)} \cap G(u))
  \end{array}\]
\end{definition}
The two definitions of the termination set $\pre(F\dtl G)$ are equivalents;
it can be proved by distribution of  union over intersection. In another hand
we remember that $\grs(F) = \cpl{F(\ets)}$. 

The set transformer $(F\dtl G)(r)$, for any $r$ in $\Pow(u)$ associated
with the dovetail operator is obtained from the pairing condition
(\ref{paircond}): 
\begin{equation}
  \label{strdtl}
  (F\dtl G)(r) = \Li(F\dtl G)(r) \cap \pre(F\dtl G) 
\end{equation}

We note that as far as the liberal set transformed is concerned, the
dovetail operator is equal to the choice operator. It differs by
having a more liberal pairing condition: to ensures that $F\dtl G$
halts, it suffices to forbid $F$ and $G$ from both looping and to
forbid either from looping in a state where the other fails.

As in \cite{BrNeAFCD}, we can prove: $\grd(F\dtl G) = \grd(F) \Or
\grd(G)$, but we give a shorter proof than \cite{BrNeAFCD} in terms of
sets. We prove:
\begin{equation}
  \label{guarddtl}
  \cpl{(F\dtl G)(\ets)} = \cpl{F(\ets)} \cup \cpl{G(\ets)}
\end{equation}

\bop
\begin{Preuve}
  \cpl{F(\ets)} \cup \cpl{G(\ets)} \\
  \Com{=}{ }\\
  \cpl{F(\ets)\cap G(\ets)}\\
  \Com{=}{ Pairing Condition}\\
  \cpl{\Li(F)(\ets)\cap \Li(G)(\ets) \cap F(u)\cap G(u)} \\
  \Com{=}{ $F(\ets)\cap\cpl{F(\ets)} = \ets $}\\
  \cpl{\Li(F)(\ets)\cap \Li(G)(\ets) \cap (F(u)\cap G(u) \cup F(\ets)\cap
  \cpl{F(\ets)})} \\
  \Com{=}{ $\Li(F)(\ets)\cap F(\ets) = \Li(F)(\ets) \cap F(u)$ See
  note below }\\
  \cpl{\Li(F)(\ets)\cap \Li(G)(\ets) \cap (F(u)\cap G(u) \cup F(u)\cap
  \cpl{F(\ets)})} \\
  \Com{=}{ Similar to two last steps }\\
  \cpl{\Li(F)(\ets)\cap \Li(G)(\ets) \cap (F(u)\cap G(u) \cup F(u)\cap
  \cpl{F(\ets)} \cup G(u)\cap \cpl{G(\ets)}} ) \\
  \Com{=}{ Definition of $(F\dtl G)(\ets)$ (\ref{Ldtl}) and
  (\ref{strdtl}) }\\
  \cpl{(F\dtl G)(\ets)}
\end{Preuve}
\eop
\emph{Note\/} In \cite{BrNeAFCD} this step requires the proof of
$\m{wlp}.F.\False \Imp \grd.F = \neg \m{hlt}.F$. We denote this
implication as a set expression: $\Li(F)(\ets)\cap F(\ets) =
\Li(F)(\ets) \cap F(u)$. However the proof of this expression is
easily given by the pairing condition. We finally note that the sets
$F(\ets)$ and $F(u)$ are not equals as we can think from the given
equality; only the intersection of these sets with $\Li(F)(\ets)$
is equal.

 The dovetail operator is in general non monotonic for the
approximation order in commands as defined in \cite{BrNeAFCD}.  Therefore
the existence of least fixed points of recursive equations cannot be proved
generally. However, the existence of least fixed points in a restricted
class of recursive definitions containing the dovetail operator, is proved
in \cite{BrNeAFCD}. In this report we only use the dovetail operator to
model fair iteration of events. We do not propose the use of this
operator to model or refine \B\ event systems. The set transformer
modeling fair iteration of events with the dovetail operator is
monotonic in the set inclusion order. 

\section{Basic Liveness Properties}\label{BLP}

Let $S$ be a \B\ event system with state variable $x$ and invariant $I$,
made up of a family of events indexed by a certain index set $L$:
\[S \Def \;\Sel_{i\in L} F_{i}\] where $\Sel_{i\in L} F_i$ denotes the
choice of events $F_i$ over a set $L$. Since we cannot ensure the
execution of continuously enabled events with an infinite set of
events in a system, as required by the weak fairness assumption, the
set of labels $L$ must be finite. Let $P$ and $Q$ be two predicates on
the state of $S$. A basic liveness property, specified by the relation
\emph{ensures\/} ($\gg_w$) as:
\[G\cdot P \gg_{w} Q\]
(pronounce ``by event $G$, $P$ ensures $Q$''), where $G =\;
\Sel_{i\in K}F_{i}$ and $K$ is a non empty subset of $L$, indicates
that by the execution of event $G$ in a state where the state variable
$x$ satisfies $P$, the system goes to another state where the
state variable satisfies $Q$, under a weak fairness assumption.

The sufficient conditions to guarantee that system $S$ satisfies the
property $G\cdot P \gg_{w} Q$ are:
\begin{center}
\begin{tabular}{|c|l|c|}\hline
&\ ANTECEDENT &~~~CONSEQUENT~~~\\ \hline
& & \\[-2mm]
\textbf{~~~WF0~~~}&\ $I\And P\And\neg Q\Imp [S]\,P\Or Q$~~~~& %
   \ $G\cdot P\gg_{w} Q $
\\[1mm]
\textbf{~WF1~} & \  $I\And P \And \neg Q \Imp \grd(G)\And [G]\,Q$~~~ & 
\\[2mm]
\hline
\end{tabular}
\end{center}

If we consider the choice of events which does not establish
postcondition $Q$, we can restate the proof obligations as follows:
\begin{center}
\begin{tabular}{|c|l|c|}\hline
&\ ANTECEDENT &~~~CONSEQUENT~~~\\ \hline
& & \\[-2mm]
\textbf{~~~WF0'~~~}&\ $I\And P\And\neg Q\Imp [F]\,P\Or Q$~~~~& %
   \ $G\cdot P\gg_{w} Q $
\\[1mm]
\textbf{~WF1'~} & \  $I\And P \And \neg Q \Imp \grd(G)\And [G]\,Q$~~~ & 
\\[2mm]
\hline
\end{tabular}
\end{center}
where $F=\;\Sel_{i\in L-K}F_{i}$. As we have $S = F\Sel G$, we can
prove the equivalence between the antecedents of WF0 and WF1 with WF0'
and WF1'.

In the following section we proof that WF0 and WF1 are indeed
sufficient conditions to guarantee that by the execution of event $G$
in a state satisfying $P$, the system goes to another state satisfying
$Q$, under a weak fairness assumption. However, as we prove our rules
in a set theoretical framework, we give an equivalent definition of
proof obligations WF0 and WF1 in term of set transformers.  In this
way, each event $F_i$ in \B\ system $S$, is considered as a set
transformer of type $\Pow(u) \Totfunc \Pow(u)$, where $u = \{ z \;|\;
I \}$ is the set of states satisfying invariant $I$. According to
\cite{AbrTBB}, the set transformer \textsf{str}($F_i$) is defined as
follows:
\[\textsf{str}(F_{i}) = \lambda r\cdot (r\in\Pow(u)\;|\; \{ z \;|\; z\in u \And
[F_{i}]\;z\in r \}) \] Following the notation introduced in
\cite{AbrMusIDCB}, we use names of events to denote set transformers.
In this way $F_{i}(r)$ denotes the largest subset of $u$, where the
execution of event $F_i$ must start in order to terminate in a state
belonging to $r$. Now, considering the sets:
\begin{eqnarray*}
  p & = & \{z\;|\;z\in u \And P \} \\
  q & = & \{z\;|\;z\in u \And Q \}
\end{eqnarray*}
the inclusions
\begin{eqnarray}
  \label{wf0s}
  p\cap\cpl{q} & \subseteq & S(p\cup q) \\
  \label{wf1s}
  p\cap\cpl{q} & \subseteq & \grs(G) \cap G(q)
\end{eqnarray}
are equivalent to WF0 and WF1 respectively. To prove the equivalences
we assume that $I\Imp [S]\;I$ holds. Then we have:

\bop
\begin{Preuve}
\m{WF0}\\
\Com{\Imp}{ Def. of WF0 and assumption }\\
\forall x\cdot (I\And P\And \neg Q\Imp [S]\,(P\Or Q))\And \forall x\cdot (I\Imp [S]\,I)\\
\Com{\Imp}{ $S$ is conjunctive }\\
\forall x\cdot (I\And P\And \neg Q\Imp [S]\,((P\Or Q)\And I))\\
\Com{\equiv}{ def. $p$, $q$ and set. trans. }\\
\forall x\cdot (x\in p\cap \cpl{q}\Imp x\in S(p\cup q))\\
\Com{\equiv}{ }\\
p\cap \cpl{q}\subseteq S(p\cup q)\\
\Com{\equiv}{ }\\
\forall x\cdot (I\And P\And \neg Q\Imp [S]\,((P\Or Q)\And I))\\
\Com{\Imp}{ weakening }\\
\m{WF0}
\end{Preuve}
\begin{Preuve}
\m{WF1}\\
\Com{\Imp}{ Def. of WF1 and assumption }\\
\forall x\cdot (I\And P\And \neg Q\Imp \grd(G)\And [G]\,Q)\And \forall
x\cdot (I\Imp [S]\,I)\\ 
\Com{\Imp}{ $G$ is conjunctive }\\
\forall x\cdot (I\And P\And \neg Q\Imp \grd(G)\And [G]\,(Q\And I))\\
\Com{\equiv}{ def. $p$, $q$ }\\
\forall x\cdot (x\in p\cap \cpl{q}\Imp \neg ([G]\,\False)\And [G]\,x\in q)\\
\Com{\equiv}{ Def. set. trans. }\\
\forall x\cdot (x\in p\cap \cpl{q}\Imp x\in \cpl{G(\ets)}\And x\in G(q))\\
\Com{\equiv}{ }\\
p\cap \cpl{q}\subseteq \grs(G)\cap G(q)\\
\Com{\Imp}{ Weakening }\\
\m{WF1}
\end{Preuve}

\eop

\subsection{Termination of Fair Iteration}

The general strategy in the proof of a basic liveness properties $P
\gg_{w} Q$ is to divide the events of $S$ into two groups: one for the
events that establish $Q$ and another one for the events that maintain
$P$ or establish $Q$. The first group is characterized by event $G$,
and the second one by an event $F$, where $F=\;\Sel_{i\in L-K}F_{i}$.
Events $F$ and $G$ are modeled by conjunctive set transformers of type
$\Pow(u)\Totfunc\Pow(u)$, and the \B\ event system $S$ can be seen as:
\begin{equation}
  \label{Asys}
  S \Def F \Sel G 
\end{equation}

As we know, most of the time, an abstract system like $S$ does not
terminate. For this reason we cannot speak about the establishment of a
certain postcondition $Q$ when $S$ terminates. In \cite{AbrMusIDCB}, this
situation is managed by translating the problem of reachability of a
certain postcondition $Q$ in a system $S$ to the problem of termination of
the iteration $(\neg Q \Rarrow S)\open$. We follow a similar approach, but
we consider a fair  iteration with the help of the dovetail operator.


\noindent Let $q$ be a subset of $u$ and $X(q)$ be the following iteration:
\begin{equation}
  \label{X}
  X(q) = \cpl{q}\Rarrow ((F\sco X(q)) \dtl G) 
\end{equation}
Since all events in $S$ always terminate, we conclude that $F$ and $G$
always terminate.  Therefore we expect that $X(q)$ eventually
terminates when it is executed in any state of $\cpl{G(\ets)}\cap
\cpl{q}$.  This expectation is ensured with the semantic of the
dovetail operator, which guarantees that $G$ will be eventually
executed. On the other hand, if $X(q)$ starts its execution in any state
of $q$, the guard of $X(q)$ is not enabled and the state of the system
is not changed. This is formally stated in the following lemma:

\begin{lemma}
    \label{termination}
    {\rm (Termination)}\\
    Let $X(q)$ be a fair iteration, $X(q) = \cpl{q}\Rarrow (F\sco X)
    \dtl G$, where $F$ and $G$ are conjunctive set transformers of
    type $\Pow(u)\Totfunc\Pow(u)$, $\pre(F)=u$ and $\pre(G)=u$. Then
    the inclusion $\grs(G)\cup q\subseteq \pre(X(q))$ holds.
\end{lemma}

\noindent\bop
\begin{Preuve}
\cpl{G(\ets)}\cup q\\
\Com{=}{ $G(u)=u$ }\\
\cpl{G(\ets)}\cap G(u)\cup q\\
\Com{\subseteq}{ $Z= (F\sco X(q))$  }\\
Z(u)\cup \cpl{G(\ets)}\cap G(u)\cup q\\
\Com{=}{ absorption }\\
Z(u)\cup (Z(u)\cap \cpl{Z(\ets)})\cup (\cpl{G(\ets)}\cap G(u))\cup q\\
\Com{=}{ def. dovetail (\ref{Ldtl}), $Z$ and $G(u)=u$ }\\
\pre((F\sco X(q))\dtl G)\cup q\\
\Com{=}{ def. termination set  }\\
((F\sco X(q))\dtl G)(u)\cup q\\
\Com{=}{ def. set transformer }\\
(\cpl{q}\Rarrow ((F\sco X(q))\dtl G))(u)\\
\Com{=}{ def. termination set and (\ref{X})}\\
\pre(X(q))
\end{Preuve}
\eop

\subsection{Total Correctness of Fair Iteration}\mbox{}\\

\noindent From lemma \ref{termination}, we assert that the fair
iteration $X(q)$ always terminates when it is executed in a state
where $\grd(G)$ holds. Informally, this fact results from the
operational meaning of the dovetail operator. As it was indicated in
section \ref{dovetail}, the two operands in the dovetail operator,
$F\sco X(q)$ and $G$, are executed fairly in parallel on separate
copies of the state, accepting as an outcome any proper, nonlooping
outcome of either operand. If $F$ does not preserve $\grd(G)$, the
sequence $F\sco X(q)$ may loop forever depending on the guard of $F$;
however in this case the semantic of the dovetail operator guarantees
that $X(q)$ terminate because $G$ do so when it is executed in a state
of $\grs(G)$ . This behavior is not exactly the same as in the \B\ event
system \m{S} because of the guards: event $G$ cannot be executed in
a state where $\neg \grd(G)$ holds. In order to improve our model of
fair iteration among events, we add the constraint that $F$ must
preserve the guard of $G$ . In this way, the behaviors of $S$ under
the weak fairness assumption and $X(q)$ are similar. Furthermore, if
$G$ is able to establish $q$ when it starts its execution in a state
in $p\cap\cpl{q}$, for a certain subset $p$ of $u$, if $p\cap\cpl{q}$
is a subset of $\grs(G)$ and $F$ preserves $p$ or establishes $q$ when
it is executed in any state of $p\cap\cpl{q}$, then we can assert that
$X(q)$ terminates in a state of $q$ when it is executed in any state
of $p\cap\cpl{q}$.  This reasoning is formalized in the following
lemma:

\begin{lemma}\label{totcor}
  {\rm (Total Correctness)}\\
  Under assumptions of lemma \ref{termination}, and for any $p$ and
  $q$ in $\Pow(u)$, such that $p\cap\cpl{q}\subseteq F(p\cup q)$,
  $p\cap\cpl{q}\subseteq~\grs(G)$, and $p\cap\cpl{q} \subseteq G(q)$
  then $p\cup q\subseteq X(q)(q)$ holds.
\end{lemma}

\noindent\bop\\
According to the pairing condition (\ref{strdtl}), the goal of lemma
\ref{totcor} becomes:
\begin{eqnarray}
\label{totcorOP1}
\lefteqn{p\cup q\subseteq \Li(X(q))(q)} \\
\label{totcorOP2}
\lefteqn{p\cup q\subseteq \pre(X(q))}
\end{eqnarray}


In order to prove subgoal (\ref{totcorOP1}), we note the following
equality for any $r$ in $\Pow(u)$:
\begin{equation}
  \label{LiXqR}
  \Li(X(q))(r) = \Cf(q)(r)(\Li(X(q))(r))
\end{equation}
where $\Cf(q)(r)$, for any subset $q$ and $r$ of $u$, is the set transformer:
\begin{eqnarray}
\label{Cf3}
\lefteqn{\Cf(q)(r)=\cpl{q}\Rarrow (G(r)\; |\; F)}
\end{eqnarray}
Equality (\ref{LiXqR}) is proved as follows:\\
\begin{Preuve}
\Li(X(q))(r)\\
\Com{=}{ (\ref{X}) }\\
\Li(\cpl{q}\Rarrow ((F\sco X(q))\dtl G))(r)\\
\Com{=}{ def. Liberal of guard }\\
q\cup \Li((F\sco X(q))\dtl G)(r)\\
\Com{=}{ def. of dovetail \ref{Ldtl} }\\
q\cup \Li(F\sco X(q))(r)\cap \Li(G)(r)\\
\Com{=}{ $G(u)=u$ and property (\ref{LieqStr}) }\\
q\cup \Li(F\sco X(q))(r)\cap G(r)\\
\Com{=}{ def. Liberal of sequencing }\\
q\cup \Li(F)(\Li(X(q))(r))\cap G(r)\\
\Com{=}{ $F(u)=u$ and property (\ref{LieqStr}) }\\
q\cup F(\Li(X(q))(r))\cap G(r)\\
\Com{=}{ def. preconditioned set transformer }\\
q\cup (G(r)\; |\; F)(\Li(X(q))(r))\\
\Com{=}{ def. guarded set transformer }\\
(\cpl{q}\Rarrow (G(r)\; |\; F))(\Li(X(q))(r))\\
\Com{=}{ def. of $\Cf(q)(r)$ (\ref{Cf3}) }\\
\Cf(q)(r)(\Li(X(q))(r))
\end{Preuve}

We note that $\Cf(q)(r)$ is a monotonic set transformer, that is for
any subset $s$ and $t$ of $u$, such that $s\subseteq t$ we have:
\begin{Preuve}
s\subseteq t\\
\Com{\Imp}{ monotonic $F$ }\\
F(s)\subseteq F(t)\\
\Com{\Imp}{  }\\
G(r)\cap F(s)\subseteq G(r)\cap F(t)\\
\Com{\Imp}{  }\\
q\cup (G(r)\cap F(s))\subseteq q\cup (G(r)\cap F(t))\\
\Com{\Imp}{ def. set transformer }\\
(\cpl{q}\Rarrow (G(r)\; |\; F))(s)\subseteq (\cpl{q}\Rarrow (G(r)\; |\; F))(t)\\
\Com{=}{ (\ref{Cf3}) }\\
\Cf(q)(r)(s)\subseteq \Cf(q)(r)(t)
\end{Preuve}
Therefore, as indicated in section \ref{Lib_set}, a recursive
definition of a liberal set transformer
$\Li(X(q))(r)=\Cf(q)(r)(\Li(X(q))(r))$, with monotonic $\Cf(q)(r)$, allow us
to state:
\begin{equation}
\label{LiXqFIX}
\Li(X(q))(r)=\FIX(\Cf(q)(r))
\end{equation}
Furthermore, we note
\begin{equation}
  \label{LiXq}
  \FIX(\Cf(q)(r))=\bigcup \Phi_{r}^{q} \quad 
\end{equation}
where
$\Phi_{r}^{q} = \{ x \,|\, x\in\Pow(u) \And x\subseteq \Cf(q)(r)(x)\}$.

Finally, the proof of  subgoal \ref{totcorOP1} is as follows:
\begin{PreuveL}{ }
1 & p\cap \cpl{q}\subseteq G(q)\cap F(p\cup q)\cup q &  From Hyp.\\
2 & p\cap \cpl{q}\subseteq (\cpl{q}\Rarrow (G(q)\; |\; F))(p\cup q) &
    1 and set trans. \\ 
3 & p\cap q\subseteq G(q)\cap F(p\cup q)\cup q &  Trivial\\
4 & p\subseteq (\cpl{q}\Rarrow (G(q)\; |\; F))(p\cup q) &  3 and 2\\
5 & q\subseteq q\cup G(q)\cap F(p\cup q) &  trivial\\
6 & q\subseteq (\cpl{q}\Rarrow (G(q)\; |\; F))(p\cup q) &  5 and set trans.\\
7 & p\cup q\in (\cpl{q}\Rarrow (G(q)\; |\; F))(p\cup q) &  6 and 4\\
8 & p\cup q\in \Cf(q)(q)(p\cup q) & 7 and (\ref{Cf3}) \\
9 & p\cup q\in \Phi_{q}^{q} &  8 and def. $\Phi_{q}^{q}$\\
10 & p\cup q\subseteq \bigcup \Phi_{q}^{q} &  9\\
11 & p\cup q\subseteq \FIX(\Cf(q)(q)) & 10 and (\ref{LiXq})\\
12 & p\cup q\subseteq \Li(X(q))(q) &  11 and (\ref{LiXqFIX}) 
\end{PreuveL}
The proof of subgoal \ref{totcorOP2} which terminates the proof of
lemma \ref{totcor} is:
\begin{PreuveL}{ }
1 & p\cap \cpl{q}\subseteq \cpl{\m{G'}(\ets)} &  Hyp.\\
2 & p\cap q\subseteq q &  trivial\\
3 & p\subseteq \cpl{\m{G'}(\ets)}\cup q &  2 and 1\\
4 & p\subseteq \pre(X(q)) &  3 and lemma \ref{termination}\\
5 & q\subseteq \cpl{\m{G'}(\ets)}\cup q &  trivial\\
6 & q\subseteq \pre(X(q)) &  5 and lemma \ref{termination}\\
7 & p\cup q\subseteq \pre(X(q)) &  6 and 3 
\end{PreuveL}
\eop

As we can see, the hypothesis in lemma (\ref{totcor}):
$p\cap\cpl{q}\subseteq (F\Sel G)(p\cup q)$, $p\cap\cpl{q} \subseteq
\cpl{G(\cpl{\ets})}$ and $p\cap\cpl{q} \subseteq G(q)$ are the
corresponding proof obligations (\ref{wf0s}) and (\ref{wf1s}) for a
basic liveness property $G\cdot P\gg_{w} Q$ of system $S$.
These inclusions, the implicit assumption that all events in $S$
always terminate and the fairness assumption, are the guarantee that
iteration of events in $S$, starting at any state in $P\And\neg Q$
will certainly terminate in a state into $Q$, which is the intended
meaning of the basic liveness property.

\subsection{Guard of the Fair Loop}

As we know, for any monotonic set transformer $S$, the complement of
the guard of $S$, $S(\ets)$, denotes the set of states where the
execution of $S$ is impossible. In the other hand, $S$ becomes a
miraculous statement when its execution ``starts'' in any state of
$S(\ets)$, and it is able to establishes any postcondition $q$,
because $S$ is monotonic and then $S(\ets)\subseteq S(q)$ holds for
any subset $q$ of $\dom(S)$.

Before we calculate the guard of the fair loop, we prove the following
lemma indicating that $X(q)$ is a monotonic set transformer:

\begin{lemma}\label{monotony}
  {\rm (Monotony of the Fair Loop)}\\
  For any subset $s$ and $t$ of $u$, such that $s\subseteq t$ we have
  $X(q)(s)\subseteq X(q)(t)$
\end{lemma}
\noindent\bop
\begin{Preuve}
s\subseteq t\\
\Com{\Imp}{ Monotony of $G$ }\\
G(s)\subseteq G(t)\\
\Com{\Imp}{ Fact of sets for any $y\in\Pow(u)$ }\\
q\cup (G(s)\cap F(y))\subseteq q\cup (G(t)\cap F(y))\\
\Com{\equiv}{ def. $\Cf(q)(s)$ and $\Cf(q)(t)$ }\\
\Cf(q)(s)(y)\subseteq \Cf(q)(t)(y)\\
\Com{\Imp}{ Fact of sets for any $y\in\Pow(u)$ }\\
\{\,y\,|\,y\subseteq u\And y\subseteq \Cf(q)(s)(y)\,\}\subseteq
\{\,y\,|\,y\subseteq u\And y\subseteq \Cf(q)(t)(y)\,\}\\ 
\Com{\Imp}{ from (\ref{LiXq}) }\\
\FIX(\Cf(q)(s))\subseteq \FIX(\Cf(q)(t))\\
\Com{\equiv}{ from (\ref{LiXqFIX}) }\\
\Li(X(q))(s)\subseteq \Li(X(q))(t)\\
\Com{\Imp}{ }\\
\Li(X(q))(s)\cap \pre(X(q))\subseteq \Li(X(q))(t)\cap \pre(X(q))\\
\Com{\equiv}{ Pairing condition (\ref{paircond}) }\\
X(q)(s)\subseteq X(q)(t)
\end{Preuve}
\vspace{-7mm}\eop

Now we state the following lemma:
\begin{lemma}\label{guardX(q)}
{\rm (Guard of Fair Loop)}\\
The guard of $X(q)$ is the complement of the least fixpoint of
$\Cf(q)(\ets)$ ($\cpl{X(q)(\ets)}=\cpl{\fix(\Cf(q)(\ets))})$)
\end{lemma}
In order to prove lemma \ref{guardX(q)}, we prove
\[ X(q)(\ets)=\fix(\Cf(q)(\ets)) \]
as follows:\\
\bop
\begin{Preuve}
X(q)(\ets)\\
\Com{=}{ Def. $X(q)$ }\\
q\cup ((F\sco X(q))\dtl G)(\ets)\\
\Com{=}{ From (\ref{guarddtl}) }\\
q\cup ((F\sco X(q))(\ets)\cap G(\ets))\\
\Com{=}{ Set Transformers }\\
(\cpl{q}\Rarrow (G(\ets)\; |\; F))(X(q)(\ets))\\
\Com{=}{ Def. $\Cf(q)(\ets)$ (\ref{Cf3}) }\\
\Cf(q)(\ets)(X(q)(\ets))\\
\Com{=}{ $\Cf(q)(\ets)$ is a monotonic function }\\
\fix(\Cf(q)(\ets))
\end{Preuve}
\eop

As the complement of the guard of $X(q)$ is the least fixpoint of
$\Cf(q)(\ets)$, we know that $\fix(\Cf(q)(\ets))$ contains all finite
chains terminating out of the guard of $\Cf(q)(\ets)$, that is, in the
set $q\cup (G(\ets)\cap F(\ets))$. Formally, this fact is stated as
follows: 
\begin{equation}
  \label{chinfix}
  \forall i\cdot (i\in\Nat \Imp
  \Cf(q)(\ets)^{i}(\Cf(q)(\ets)(\ets))\subseteq \fix(\Cf(q)(\ets)) 
\end{equation}
\bop\\
Let $r$ be any set in $\{ z \;|\; z\subseteq u \And
\Cf(q)(\ets)(z)\subseteq z \}$. We prove by induction:
\begin{equation}
  \label{chinfix1}
  \forall i\cdot (i\in\Nat \Imp \Cf(q)(\ets)^{i+1}(\ets)
  \subseteq r) 
\end{equation}
Base Case:
\begin{Preuve}
\Cf(q)(\ets)^{0+1}(\ets)\\
\Com{=}{ }\\
\Cf(q)(\ets)(\ets)\\
\Com{\subseteq}{ Monotony of $\Cf(q)(\ets)$ }\\
\Cf(q)(\ets)(r)\\
\Com{\subseteq}{ Hyp. $\Cf(q)(\ets)(r)\subseteq r$ }\\
r
\end{Preuve}
Inductive Step:\\
\begin{PreuveL}{ }
1 & \Cf(q)(\ets)^{i+1}(\ets)\subseteq r&  Ind. Hyp.\\
2 & \Cf(q)(\ets)(\Cf(q)(\ets)^{i+1}(\ets))\subseteq \Cf(q)(\ets)(r) &
1, Mon. of $\Cf(q)(\ets)$\\ 
3 & \Cf(q)(\ets)^{i+2}(\ets)\subseteq r &  2 and Hyp.
\end{PreuveL}
Now, (\ref{chinfix}) follows from (\ref{chinfix1}), considering that
$\fix(\Cf(q)(\ets))$ is the generalized intersection of all subsets in
$\{ z \;|\; z\subseteq u \And \Cf(q)(\ets)(z)\subseteq z \}$.
\eop

From monotony of $X(q)$ (lemma \ref{monotony}), the guard of
$X(q)$ (lemma \ref{guardX(q)}) and (\ref{chinfix}), it follows, for
any $r\in\Pow(u)$:
\[ \forall i\cdot (i\in\Nat \Imp \Cf(q)(\ets)^{i+1}(\ets)
  \subseteq X(q)(r)) \]
This last inclusion indicates that the set of states that guarantees
termination of $X(q)$ in any state of $r$, contains all states where
any iteration of $(\cpl{q}\Rarrow G(\ets)\;|\;F)$ terminates in $q
\cup (G(\ets)\cap F(\ets))$. Moreover, if $X(q)$ starts execution in
any state of $\Cf(q)(\ets)^{i+1}(\ets)$ for any $i\in\Nat$, $X(q)$
becomes a miraculous statement, able to establish any postcondition.

\section{General Liveness Properties}\label{GLP}

In \B\ event system $S$, with state variable $x$ and invariant $X$,
general liveness properties are specified by formulae $P \leadsto Q$,
where $P$ and $Q$ are predicates on the system state. This property
specifies that the system eventually reaches a state satisfying $Q$
whenever it reaches any state in $P$. There are three basic
differences between a $\leadsto$ relation and a $\gg_{w}$ relation.
The first difference is the number of steps involved in the transition
from $P$ to $Q$. With $\gg_{w}$, the helpful transition is done by the
execution of an atomic event, while with $\leadsto$, the number of
atomic transitions is not specified. The second difference is that we
can assert with $G \cdot P \gg_{w} Q$ that the system maintains $P$
while $Q$ is not established.  We do not have this guarantee when we
specify $P \leadsto Q$. Finally, the third difference is that a
general liveness property does not directly depend on any fairness
assumption while a basic liveness property do.

A property $P\leadsto Q$ holds in a B event system if it is derived by
a finite number of applications of the rules defined by the {\sc
  unity} theory:
\begin{center}
\begin{tabular}{|c|l|l|}\hline
 &\ ANTECEDENT &\ ~~CONSEQUENT \\ \hline
& & \\[-2mm]
\textbf{~~BRL~~~}&\ $G\cdot P \gg_{w} Q$ &\ $P \leadsto Q$ \\[1mm]
\textbf{~TRA~} &\ $P\leadsto R$, $R\leadsto Q$ &\ $P \leadsto Q$ \\[1mm]
\textbf{~DSJ~} &\ $\forall m \cdot (m\in M \Imp P(m)\leadsto Q)~~$&\  $\exists
m \cdot (m\in M \And P(m))\leadsto Q$~~~~ \\[2mm]
\hline
\end{tabular}
\end{center}

So as to reason about liveness properties, we incorporate the proof
system in {\sc unity} in the framework of \B\ event systems. We can
use all theorems in \cite{ChMiPPD} concerning \emph{ensures} and
\emph{leads to\/} relations in the rules of proof of several
properties of \B\ event systems.  

\subsection{Refining Liveness Properties}\label{PORef}

If abstract system $S$ is refined into another one $T$ we need to
assert that any abstract property $\PP$ is preserved in $T$. As property
$\PP$ depends on basic properties $\QQ$, we only need to demonstrate that
each basic property $\QQ$ is preserved in $T$. We can establish the
validity of each property $\QQ$ in the refinement $T$ by the proof of WF0
and WF1 proof obligations. However, if we do these proofs, we would repeat
the proofs done in the abstraction $S$ because WF0 is completely preserved
by refinement and WF1 is partially preserved. So as to reduce the number
and complexity of proofs, we propose two new proof obligations that the
refinement $T$ must satisfy in order to preserve a basic liveness property.
We present these proof obligations for a certain basic liveness property
$\QQ$.

Let $\QQ$ be the property $G\cdot P \gg_{w} Q$ which holds in abstract
system $S$. From WF0' and WF1' in section \ref{BLP}, we know that $S$ can be
considered as an event system $F\Sel G$, where $F = \Sel_{i\in L-K}F_{i}$,
such that $P\And \neg Q \Imp [F]\,(P\Or Q)$ and $P\And \neg Q\Imp [G]\,Q$
holds under the assumption of $I$. If $S$ is refined to $T$, the refinement
is considered as an event system $F'\Sel G' \Sel H$, where $F'$ and $G'$
are the refinements of $F$ and $G$ respectively and $H$ are new events that
refine $\Skip$ \cite{AbrMusIDCB}. We consider that the abstract state is
refined by a concrete one, and these states are related by the gluing
invariant $J$.  Under the assumptions $I\And J$, and according to the very
definition of refinement, we conclude that $P\And \neg Q\Imp [F'\Sel
H]\,(P\Or Q)$ and $P\And \neg Q\Imp [G']\,Q$ hold in $T$. However we cannot
assert that $P\And \neg Q\Imp \grd(G')$ holds under the same assumptions,
because the refined event $G'$ has a guard stronger than $\grd(G)$.  Then,
in order to guarantee the preservation of $\QQ$ we need to prove that $T$
reaches a state in the guard of $G'$ when it is in a state out of the guard
(rule LIP: Liveness Preservation), and that the guard of $G'$ is preserved
by $F'$ and $H$ (rule SAP: Safety Preservation).  Formally the proof
obligations in $T$ are:
\begin{center}
\begin{tabular}{|c|l|}\hline
& \\[-2mm]
\textbf{~~~LIP~~~} &\ $I\And J\And P\And \neg
Q\And\neg\grd(G')\leadsto\grd(G')$\\[1mm] 
\textbf{~SAP~} &\ $I\And J\And P\And\neg Q\And\grd(G')\Imp [F'\Sel
H]\,grd(G')$~~~~\\[2mm] 
\hline
\end{tabular}
\end{center}

We conclude this section by a summary of our approach to the
specification and refinement of a general liveness property $\PP$.
Property $\PP$ is proved in the abstract system $S$ by identifying
basic liveness properties $\QQ$, such that $\PP$ is derived from $\QQ$
by application of rules given in section \ref{GLP}. Each property
$\QQ$ is then proved by WF0 and WF1 proof obligations. When system $S$
is refined to system $T$, each property $\QQ$ in $S$ generates new
proof obligations $\PP'_1$ and $\PP'_2$ in $T$ as stated by LIP and
SAP rules. In turn, in order to prove each general liveness property
$\PP'_1$, we need to identify other basic liveness properties $\QQ'$.
We continue this process at each step of refinement. We observe that
properties $\QQ$ and $\QQ'$ specify an atomic transition at each step
of refinement.  However, the transition at step $i+1$ of a refinement
is ``shorter'' than the transition at step $i$. That is, at level $i$
a certain basic property $\QQ$ specifies an atomic transition from a
state in $P$ to another one in $Q$. At level $i+1$ we do not need to
prove the (concrete) transition from $P$ to $Q$, we are only concerned
with the transition specified in $\QQ'$ which is necessary in the
proof of the transition from a state in $r^{-1}[p\cap
\cpl{q}]\cap{\cpl{\grs(G')}}$ to another one in $\grs(G')$, where $G'$
is the refinement of the helpful event related to $\QQ$. In this way,
our method of specification and proof of liveness properties becomes a
guide that serves to specify and prove the dynamic behavior of a
system at each level of refinement.

In the next subsection we give a justification of proof obligations
LIP and SAP, as sufficient conditions to ensures the preservation of
liveness properties under refinement.

\subsection{Proving Refinement of Basic Liveness Properties under Weak Fairness}
\label{RBLP}

When abstract system $S$ (\ref{Asys}) is refined, the abstract events $F$
and $G$ are refined by concrete events $F'$ and $G'$ respectively and new
events $H$ appear. In this way, the abstract system $S$ is refined by the
system $S'$:
\begin{equation}
  \label{Csys}
  S' \Def F' \Sel G' \Sel H 
\end{equation}
Let $y$ be the concrete state variable of $S'$ and $v$ the concrete
state space, where $v = \{ y \;|\; \exists x \cdot ( I(x) \And J(x,y))
\}$, $I$ is the abstract invariant of $S$ and $J$ the gluing invariant
of $T$.  The events $F'$, $G'$ and $H$ are modeled by conjunctive set
transformers of type $\Pow(v)\Totfunc\Pow(v)$. The abstract and
concrete events are related by the refinement relation: $F\sqsubseteq
F'$ and $G\sqsubseteq G'$ and new events refine $\Skip$:
$\Skip\sqsubseteq H$. These relations among events are defined by the
following proof obligations \cite{AbrTBB}:
\begin{eqnarray}
  \label{RefF}
  \lefteqn{F(\cpl{r[\cpl{s}]})\subseteq\cpl{r[\cpl{F'(s)}]}}\\
  \label{RefG}
  \lefteqn{G(\cpl{r[\cpl{s}]})\subseteq\cpl{r[\cpl{G'(s)}]}}\\
  \label{RefS}
  \lefteqn{\Skip(\cpl{r[\cpl{s}]})\subseteq\cpl{r[\cpl{H(s)}]}}
\end{eqnarray}
where $s$ is universally quantified over $\Pow(v)$ and $r$ is a total
relation from $v$ to $u$ defined as follows $r = \{ y\mapsto x \;|\; I(x)
\And J(x,y) \}$.

From conditions stated in lemma \ref{totcor}, we know that abstract
system $S$ eventually reaches a state in $q$ when its execution
arrives at any state of $p$. In order to preserve this abstract
transition, we need to observe a concrete transition from a state in
$p'$ to another one in $q'$, where $p'$ and $q'$ are the corresponding
concrete states $r^{-1}[p]$ and $r^{-1}[q]$ respectively. In the
following paragraphs we analyze sufficient conditions for this
concrete transition.

We consider the fair iteration $X'(q')$ made up of events in $S'$:
\begin{equation}
  \label{X'}
  X'(q') = \cpl{q'}\Rarrow (((F'\Sel H)\sco X'(q')) \dtl G')
\end{equation}
This recursion models the iteration of events $F'$ and $H$ in the concrete
system. Now we state the following lemma:

\begin{lemma}\label{parcor}
  {\rm (Partial Correctness)} \\
  Under the assumptions of lemma \ref{totcor} and refinement conditions
  (\ref{RefF}), (\ref{RefG}) and (\ref{RefS}), the  inclusion
  $p'\cup q'\subseteq \Li(X'(q'))(q')$ holds.
\end{lemma}

\noindent\bop\\
A brief outline of the proof is as follows. From assumptions of 
lemma~\ref{totcor}: $F(u)=u$, $G(u)=u$, $p\cap\cpl{q}\subseteq F(p\cup q)$ and
$p\cap\cpl{q}\subseteq G(q)$ and refinement conditions (\ref{RefF}),
(\ref{RefG}) and (\ref{RefS}), the following inclusions follow:

\hspace{-\mathindent}\parbox{\textwidth}{\begin{tabular}{lc}
\parbox{.5\textwidth}{
\begin{eqnarray}
       \label{termf3}
       \lefteqn{\m{F'(v)} = v}\\
       \label{termg3}
       \lefteqn{\m{G'}(v) = v}\\
       \label{termh3}
       \lefteqn{\m{H}(v) = v}
\end{eqnarray}
}&
\parbox{.5\textwidth}{
    \begin{eqnarray}
       \label{stblf3}
       \lefteqn{r^{-1}[p\cap \cpl{q}]\subseteq \m{F'}(p'\cup q')}\\
       \label{stblg3}
       \lefteqn{r^{-1}[p\cap \cpl{q}]\subseteq \m{G'}(q')}\\
       \label{stblh3}
       \lefteqn{r^{-1}[p\cap \cpl{q}]\subseteq \m{H}(p'\cup q')}
     \end{eqnarray}
}
\end{tabular}}

We prove (\ref{termf3}) and
(\ref{stblf3}). The other proofs are done in a similar way. First, we prove
the following inclusion for any $s$ in $\Pow(u)$:
\begin{equation}
\label{safF3}
r^{-1}[F(s)]\subseteq \m{F'}(r^{-1}[s])
\end{equation}
The proof of this inclusion is based on equivalence 
$r[a]\subseteq b \equiv r^{-1}[\cpl{b}]\subseteq \cpl{a}$, where $a$
and $b$ are universally quantified over $\Pow(v)$ and $\Pow(u)$
respectively. The reference to this equivalence in the proof is given
as ``(Equ)''. The proof of (\ref{safF3}) is:
\begin{PreuveL}{ }
1 & F(\cpl{r[\cpl{r^{-1}[s]}]})\subseteq
\cpl{r[\cpl{\m{F'}(r^{-1}[s])}]} &  From (\ref{RefF})\\ 
2 & r[\cpl{\m{F'}(r^{-1}[s])}]\subseteq \cpl{F(\cpl{r[\cpl{r^{-1}[s]}]})} &  1\\
3 & r^{-1}[F(\cpl{r[\cpl{r^{-1}[s]}]})]\subseteq \m{F'}(r^{-1}[s]) &  2 and Equ\\
4 & r^{-1}[s]\subseteq r^{-1}[s] &  trivial\\
5 & r[\cpl{r^{-1}[s]}]\subseteq \cpl{s} &  4 and Equ\\
6 & s\subseteq \cpl{r[\cpl{r^{-1}[s]}]} &  5\\
7 & r^{-1}[F(s)]\subseteq r^{-1}[F(\cpl{r[\cpl{r^{-1}[s]}]})] &  6 and monotony\\
8 & r^{-1}[F(s)]\subseteq \m{F'}(r^{-1}[s]) &  7 and 3
\end{PreuveL}

\noindent The proof of (\ref{termf3}) is as follows:
\begin{PreuveL}{ }
1 & \m{F'}(v)\subseteq v &  $F\in \Pow(v)\Totfunc \Pow(v)$ \\
2 & r^{-1}[F(u)]\subseteq \m{F'}(r^{-1}[u]) &  (\ref{safF3}) and $s=u$\\
3 & r^{-1}[u]\subseteq \m{F'}(r^{-1}[u]) &  2 and $F(u)=u$\\
4 & v\subseteq \m{F'}(v) &  $r$ total: $r^{-1}[u]\!=\!v$ \\
5 & \m{F'}(v)=v &  4 and 1
\end{PreuveL}

\noindent The proof of (\ref{stblf3}) is as follows:
\begin{PreuveL}{ }
1 & r^{-1}[p\cap \cpl{q}]\subseteq r^{-1}[F(p\cup q)] &  From Hyp.\\
2 & r^{-1}[F(p\cup q)]\subseteq \m{F'}(r^{-1}[p\cup q]) &  (\ref{safF3})
    and $s=p\cup q$\\ 
3 & r^{-1}[p\cap \cpl{q}]\subseteq \m{F'}(\m{p'}\cup \m{q'}) &  2, 1,
    def. $p'$ and $q'$ 
\end{PreuveL}

\noindent With inclusions (\ref{termf3})--(\ref{stblh3}) we make a
calculus similar to the proof of subgoal $p\cup
q\subseteq\Li(X(q))(q)$ of lemma \ref{totcor}. That is, we derive the
equality $\Li(X'(q'))(q') = \FIX(\cpl{q}\Rarrow (G'(q)\,|\,(F'\Sel
H)))$ in a way similar to calculation of (\ref{LiXqFIX}). Then, using
the equality $\FIX(\cpl{q}\Rarrow (G'(q')\,|\,(F'\Sel H)))=\bigcup
\Phi'$, where $\Phi' = \{ x \,|\, x\in\Pow(v) \And x\subseteq
(\cpl{q}\Rarrow (G'(q')\,|\,(F'\Sel H)))(x)\}$ we conclude $p'\cup
q'\in \Phi'$ from refinement conditions.  Finally, from $p'\cup q'\in
\Phi$ and the last two equalities we conclude the goal of lemma
(\ref{parcor}): $p'\cup q'\subseteq\Li(X'(q'))(q')$.\\
\eop

The inclusion $\m{q'}\cup \grs(\m{G'})\subseteq \pre(\m{X'}(\m{q'}))$
follows from a calculus similar to the proof of lemma \ref{termination}.
As we know in $S$, $p\cap\cpl{q}$ is included in the guard of $G$.
Unfortunately this inclusion is not preserved by refinement, because the
guard of $G'$ is stronger than the guard of $G$ ($\grs(G')\subseteq
r^{-1}[\grs(G)]$).  Therefore the set $r^{-1}[p\cap \cpl{q}]$ is not
included in the termination set of $X'(q')$. From lemma~\ref{parcor},
inclusion $\grs(\m{G'})\subseteq \pre(\m{X'}(\m{q'}))$ and pairing
condition, we conclude $p'\cap \grs(G') \subseteq X'(q')(q')$. Furthermore,
if the guard of $G'$ is preserved by $F'$ and $H$, we can assert that
 concrete system $S'$ has a transition to a state into $q'$ whenever it
arrives at any state into $p'\cap\grs(G')$. This is formally stated in the
following lemma:

\begin{lemma}\label{lemmatsap}
 Under the assumptions of lemma \ref{totcor} and refinement conditions
  (\ref{RefF}), (\ref{RefG}) and (\ref{RefS}) as well as the following
  condition: 
  \begin{equation}
  \label{tsap3}
  r^{-1}[p\cap \cpl{q}]\cap \grs(\m{G'})\subseteq (\m{F'}\Sel
  H)(\grs(\m{G'}))
  \end{equation}
  the property $\m{G'}\cdot y\in \m{p'}\And \grd(\m{G'})\gg_{w} y\in
  \m{q'}$  holds in $S'$.  
\end{lemma}

\noindent\bop\\
We apply WF0 and WF1 proof obligations in order to prove this lemma.
First, we prove the following inclusion:
\begin{equation}
\label{inc3}
\m{p'}\cap \cpl{\m{q'}}\subseteq r^{-1}[p\cap \cpl{q}]
\end{equation}
\noindent We take $y\in \m{p'}\cap\cpl{\m{q'}}$ as premise and we prove
$y\in r^{-1}[p\cap\cpl{q}]$: 
\begin{PreuveL}{N}
1 & y\in \m{p'}\cap \cpl{\m{q'}} &  ; premise\\
2 & y\in \m{p'}\And \neg (y\in \m{q'}) &  ; 1\\
3 & y\in r^{-1}[p]\And \neg (y\in r^{-1}[q]) &  ; 2 and def $p'$ and $q'$                  \\
4 & \mbox{$\exists x\cdot (x\in p\And x\mapsto y\in r^{-1})\And \neg (\exists
    x\cdot (x\in q\And x\mapsto y\in r^{-1}))$} & ; 3\\ 
5 & \mbox{$\exists x\cdot (x\in p\And x\mapsto y\in r^{-1})\And \forall x\cdot
    (x\mapsto y\in r^{-1}\Imp x\not\in q)$} & ; 4\\ 
6 & \exists x\cdot (x\in p\And x\not\in q\And x\mapsto y\in r^{-1}) &  ; 5\\
7 & y\in r^{-1}[p\cap \cpl{q}] &  ; 6
\end{PreuveL}

\noindent Proof of WF0: $y\in \m{p'}\And \grd(\m{G'})\And y\not\in \m{q'}\Imp
[\m{F'}\Sel H]\,(y\in \m{p'}\And \grd(\m{G'})\Or y\in \m{q'})$:
\begin{PreuveL}{ }
1 & r^{-1}[p\cap \cpl{q}]\subseteq \m{F'}(\m{p'}\cup \m{q'}) &  (\ref{stblf3})\\
2 & r^{-1}[p\cap \cpl{q}]\subseteq \m{H}(\m{p'}\cup \m{q'}) &  (\ref{stblh3})\\
3 & r^{-1}[p\cap \cpl{q}]\cap \grs(\m{G'})\subseteq (\m{F'}\Sel H)(\grs(\m{G'})) &  (\ref{tsap3})\\
4 & r^{-1}[p\cap \cpl{q}]\cap \grs(\m{G'})\subseteq (\m{F'}\Sel
H)(\m{p'}\cap \grs(\m{G'})\cup \m{q'}) &  3, 2 and 1\\ 
5 & \m{p'}\cap \cpl{\m{q'}}\subseteq r^{-1}[p\cap \cpl{q}] &  (\ref{inc3})\\
6 & \m{p'}\cap \grs(\m{G'})\cap \cpl{\m{q'}}\subseteq (\m{F'}\Sel
H)(\m{p'}\cap \grs(\m{G'})\cup \m{q'}) &  5 and 4\\ 
7 & y\in \m{p'}\!\And\! \grd(\m{G'})\!\And\! y\not\in \m{q'}\Imp [\m{F'}\Sel
H]\,(y\in \m{p'}\!\And\! \grd(\m{G'})\!\Or\! y\in \m{q'}) &  6, set trans. 
\end{PreuveL}

\noindent Proof of WF1: $y\in \m{p'}\And \grd(\m{G'})\And y\not\in \m{q'}\Imp
grd(G') \And [\m{G'}]\,y\in \m{q'}$:
\begin{PreuveL}{ }
1 & r^{-1}[p\cap \cpl{q}]\subseteq \m{G'}(\m{q'}) &  (\ref{stblg3}) \\
2 & \m{p'}\cap \cpl{\m{q'}}\subseteq r^{-1}[p\cap \cpl{q}] &  (\ref{inc3})\\
3 & \m{p'}\cap \cpl{\m{q'}}\subseteq \m{G'}(\m{q'}) &  2 and 1\\
4 & \m{p'}\cap \grs(\m{G'})\cap \cpl{\m{q'}}\subseteq \grs(\m{G'})\cap
\m{G'}(\m{q'}) &  3\\ 
5 & y\in \m{p'}\And \grd(\m{G'})\And y\not\in \m{q'}\Imp
\grd(\m{G'})\And [G]\,y\in \m{q'} &  4 
\end{PreuveL}
\eop

In fact, a calculus of the termination set of $X'(q')$, using the
definition of dovetail operator (\ref{Ldtl}), allows us to conclude
$\pre(X'(q')) = \fix(\cpl{\m{q'}}\cap \cpl{\grs(\m{G'})} \Rarrow
(\m{F'}\Sel H))$. According to \cite{AbrTBB}, $\pre(X'(q'))$ is the
same set as the termination set of $(\cpl{\m{q'}}\cap
\cpl{\grs(\m{G'})} \Rarrow (\m{F'}\Sel H))\open$. From the equality
between the two sets, we conclude that the iteration of $F'$ and $H$
will stop when  $S'$ arrives into a state in $\grs(G')\cup q'$.
This reasoning allow us to propose the following lemma:
\begin{lemma}\label{lemmatlip}
  Under the assumptions of lemma \ref{lemmatsap} and the condition:
     \begin{equation}
     \label{tlip}
     y\in r^{-1}[p\cap \cpl{q}]\And \neg \grd(\m{G'})\leadsto
     \grd(\m{G'}) 
     \end{equation}
  the property $y\in \m{p'}\leadsto y\in \m{q'}$  holds in \m{S'}
\end{lemma}

\noindent\bop\\
\renewcommand{\UNLESS}{\mathrel{\mbox{\sc unless}}}
\noindent The proof of lemma (\ref{lemmatlip}) requires the proof of the following
property:
\begin{equation}\label{unless}
y\in \m{p'}\cap \cpl{\m{q'}}\And \neg \grd(\m{G'})\UNLESS
  y\in \m{p'}\cap \cpl{\m{q'}}\And \grd(\m{G'})\Or y\in \m{q'} 
\end{equation}

\noindent Let \m{lhs} be the left hand side of the \emph{unless\/} property and
\m{rhs} its right hand side. The \emph{unless\/} property follows from
$\m{lhs}\And \neg \m{rhs}\Imp [\m{S'}]\,(\m{lhs}\Or \m{rhs})$. In this
case, the property follows from the following implication:
\[
y\in \m{p'}\cap \cpl{\m{q'}}\And \neg \grd(\m{G'})\Imp
  [\m{S'}]\,(y\in \m{p'}\cap \cpl{\m{q'}}\Or y\in \m{q'}) 
\]
\parbox{\textwidth}{%
\begin{PreuveL}{ }
1 & r^{-1}[p\cap \cpl{q}]\subseteq (\m{F'}\Sel H)(\m{p'}\cup \m{q'}) &
(\ref{stblf3})  and (\ref{stblh3})\\ 
2 & r^{-1}[p\cap \cpl{q}]\subseteq (\m{F'}\Sel H)(\m{p'}\cap
\cpl{\m{q'}}\cup \m{q'}) &  1 and absorption\\ 
3 & \m{p'}\cap \cpl{\m{q'}}\subseteq r^{-1}[p\cap \cpl{q}] &  (\ref{inc3})\\
4 & \m{p'}\cap \cpl{\m{q'}}\subseteq (\m{F'}\Sel H)(\m{p'}\cap
\cpl{\m{q'}}\cup \m{q'}) &  3 and 2\\ 
5 & \m{p'}\cap \cpl{\m{q'}}\cap \cpl{\grs(\m{G'})}\subseteq
\m{G'}(\m{p'}\cap \cpl{\m{q'}}\cup \m{q'}) &  def. $grd(G')$\\ 
6 & \m{p'}\cap \cpl{\m{q'}}\cap \cpl{\grs(\m{G'})}\subseteq
\m{S'}(\m{p'}\cap \cpl{\m{q'}}\cup \m{q'}) &  5 and 4\\ 
7 & y\in \m{p'}\cap \cpl{\m{q'}}\And \neg \grd(\m{G'})\Imp
[\m{S'}]\,(y\in \m{p'}\cap \cpl{\m{q'}}\Or y\in \m{q'}) &  6 
\end{PreuveL}}

\noindent The proof uses the PSP theorem:
\[\frac{{\cal P} \leadsto {\cal Q}\;\;, {\cal R} \UNLESS {\cal S}}{%
  {\cal P}\And {\cal R} \leadsto {\cal Q} \And {\cal R} \Or {\cal S}}\]
and the cancellation (CAN) theorem:
\[\frac{P\leadsto Q \Or R, R\leadsto R' }{P \leadsto Q \Or R'}\]
\begin{PreuveL}{ }
1 & y\in r^{-1}[p\cap \cpl{q}]\And \neg \grd(\m{G'})\leadsto
\grd(\m{G'}) &  (\ref{tlip})\\ 
2 & \m{p'}\cap \cpl{\m{q'}}\subseteq r^{-1}[p\cap \cpl{q}] &  (\ref{inc3})\\
3 & \m{p'}\cap \cpl{\m{q'}}\cap \cpl{\grs(\m{G'})}\subseteq
r^{-1}[p\cap \cpl{q}]\cap \cpl{\grs(\m{G'})} &  2\\ 
4 & y\in \m{p'}\cap \cpl{\m{q'}}\And \neg \grd(\m{G'})\Imp y\in
r^{-1}[p\cap \cpl{q}]\And \neg (\grd(\m{G'})) &  3\\ 
5 & y\in \m{p'}\cap \cpl{\m{q'}}\And \neg \grd(\m{G'})\leadsto y\in
r^{-1}[p\cap \cpl{q}]\And \neg (\grd(\m{G'})) &  4\\ 
6 & y\in \m{p'}\cap \cpl{\m{q'}}\And \neg \grd(\m{G'})\leadsto
\grd(\m{G'}) &  TRA 5, 1\\ 
7 & y\in \m{p'}\cap \cpl{\m{q'}}\And \neg \grd(\m{G'})\!\leadsto \!y\in
\m{p'}\cap \cpl{\m{q'}}\!\And \!\grd(\m{G'})\!\Or \!y\in \m{q'} &  6, PSP and
(\ref{unless})\\ 
8 & y\in \m{p'}\cap \cpl{\m{q'}}\And \neg \grd(\m{G'})\leadsto y\in
\m{p'}\And \grd(\m{G'})\Or y\in \m{q'} &  7\\ 
9 & y\in \m{p'}\And \grd(\m{G'})\leadsto y\in \m{q'} &  lemma
\ref{lemmatsap} and BRL\\ 
10 & y\in \m{p'}\cap \cpl{\m{q'}}\And \neg \grd(\m{G'})\leadsto y\in q
&  CAN 9 and 8\\ 
11 & y\in \m{p'}\cap \cpl{\m{q'}}\And \grd(\m{G'})\Imp y\in p\And
\grd(\m{G'}) &  trivial\\ 
12 & y\in \m{p'}\cap \cpl{\m{q'}}\And \grd(\m{G'})\leadsto y\in \m{q'}
&  11, BRL 9, TRA\\ 
13 & y\in \m{p'}\cap \cpl{\m{q'}}\leadsto y\in \m{q'} &  DSJ 12 and 10\\
14 & y\in \m{p'}\cap \m{q'}\Imp y\in \m{q'} &  trivial\\
15 & y\in \m{p'}\leadsto y\in \m{q'} &  14,BRL, DSJ 13
\end{PreuveL}
\eop

Our last step in our proofs is to demonstrate that premises
(\ref{tsap3}) and (\ref{tlip}) of theorems \ref{lemmatsap} and
\ref{lemmatlip} are equivalent to SAP and LIP proof obligations. 

As we can see, the premises (\ref{tsap3}) and (\ref{tlip}) of theorems
\ref{lemmatsap} and \ref{lemmatlip} are the set
theoretical counterpart of SAP and LIP proof obligations which are
needed to guarantee the preservation of basic liveness properties in a
refinement. 

In order to prove the equivalence between (\ref{tsap3}) and SAP rule,
we demonstrate the following equivalences:
\begin{eqnarray}
\label{E1}
\lefteqn{y\in r^{-1}[\m{p}\cap\cpl{q}]\equiv \exists x\cdot (P(x)\And \neg
  Q(x)\And I(x)\And J(y,x))}\\
\label{E2}
\lefteqn{y\in \grs(\m{G'})\equiv y\in v\And \grd(\m{G'})}\\
\label{E3}
\lefteqn{y\in (\m{F'}\Sel H)(\cpl{\m{G'}(\ets)})\equiv y\in v\And
  [\m{F'}\Sel H]\,\grd(\m{G'})} 
\end{eqnarray}

\noindent\textbf{Proof of (\ref{E1})}
\begin{Preuve}
y\in r^{-1}[\m{p}\cap\cpl{q}]\\
\Com{\equiv}{          }\\
\exists x\cdot (x\in p\cap\cpl{q}\And y\mapsto x\in r)\\
\Com{\equiv}{ def. $p$ and $q$ }\\
\exists x\cdot (P(x)\And \neg Q(x) \And x\in u \And y\mapsto x\in r)\\
\Com{\equiv}{ def. $u$ and $r$ }\\
\exists x\cdot (P(x)\And \neg Q(x) \And I(x)\And J(y,x))
\end{Preuve}
\eop

\noindent\textbf{Proof of (\ref{E2})}
\begin{Preuve}
y\in \grs(\m{G'})\\
\Com{\equiv}{ def. $\grs(\m{G'})$  }\\
y\in \cpl{\m{G'}(\ets)}\\
\Com{\equiv}{ def. $\m{G'}(\ets)$  }\\
y\in \cpl{\{\,z\,|\,z\in v\And [\m{G'}]\,z\in \ets\,\}}\\
\Com{\equiv}{ set theory }\\
y\in \{\,z\,|\,z\in v\And \neg [\m{G'}]\,z\in \ets\,\}\\
\Com{\equiv}{ set theory }\\
y\in \{\,z\,|\,z\in v\And \neg [\m{G'}]\,\False\,\}\\
\Com{\equiv}{ set theory }\\
y\in \{\,z\,|\,z\in v\And \grd(\m{G'})\,\}\\
\Com{\equiv}{            }\\
y\in v\And \grd(\m{G'})
\end{Preuve}
\eop

\noindent\textbf{Proof of (\ref{E3})}
\begin{Preuve}
y\in (\m{F'}\Sel H)(\cpl{\m{G'}(\ets)})\\
\Com{\equiv}{ def. set transformer }\\
y\in \{\,z\,|\,z\in v\And [\m{F'}\Sel H]\,z\in \cpl{\m{G'}(\ets)}\,\}\\
\Com{\equiv}{            }\\
y\in \{\,z\,|\,z\in v\And [\m{F'}\Sel H]\,(z\in v\And \grd(\m{G'}))\,\}\\
\Com{\equiv}{$[\m{F'}\Sel H]\,z\in v\equiv z\in v$ }\\
y\in \{\,z\,|\,z\in v\And [\m{F'}\Sel H]\,\grd(\m{G'})\,\}\\
\Com{\equiv}{ set theory }\\
y\in v\And [\m{F'}\Sel H]\,\grd(\m{G'})
\end{Preuve}
\eop

\noindent The equivalence between (\ref{tsap3}) and SAP rule is as
follows:\\
\bop
\begin{Preuve}
r^{-1}[p\cap \cpl{q}]\cap \grs(\m{G'})\subseteq (\m{F'}\Sel H)(\grs(\m{G'}))\\
\Com{\equiv}{ }\\
\forall y\cdot (y\in r^{-1}[p\cap \cpl{q}]\cap \grs(\m{G'})\Imp y\in (\m{F'}\Sel H)(\grs(\m{G'})))\\
\Com{\equiv}{ }\\
\forall y\cdot (y\in r^{-1}[p\cap \cpl{q}]\And y\in \grs(\m{G'})\Imp y\in (\m{F'}\Sel H)(\grs(\m{G'})))\\
\Com{\equiv}{ (\ref{E1}), (\ref{E2}) and (\ref{E3}) }\\
\forall y\cdot (\exists x\cdot (P(x)\!\And \!\neg Q(x)\!\And\! I(x)\!\And\! J(y,x))\!\And\! y\in v\!\And\! \grd(\m{G'})\Imp y\in v\!\And\! [\m{F'}\Sel H]\,\grd(\m{G'}))\\
\Com{\equiv}{ }\\
\forall y\cdot (\exists x\cdot (P(x)\And \neg Q(x)\And I(x)\And J(y,x))\And y\in v\And \grd(\m{G'})\Imp [\m{F'}\Sel H]\,\grd(\m{G'}))\\
\Com{\equiv}{ $\exists x\cdot (P(x)\And I(x)\And J(y,x))\Imp y\in v$ }\\
\forall y\cdot (\exists x\cdot (P(x)\And \neg Q(x)\And I(x)\And J(y,x))\And \grd(\m{G'})\Imp [\m{F'}\Sel H]\,\grd(\m{G'}))\\
\Com{\equiv}{ }\\
\forall (x,y)\cdot (P(x)\And \neg Q(x)\And I(x)\And J(y,x)\And \grd(\m{G'})\Imp [\m{F'}\Sel H]\,\grd(\m{G'}))
\end{Preuve}
\eop

In order to prove the equivalence between (\ref{tlip}) and LIP proof
obligation, we need the following theorem about \emph{leads to}
\begin{equation}\label{ThLTO}
\frac{%
  (\exists x \cdot (P(x)) \And Q) \leadsto R \;,\;x\backslash
  Q\;,\;x\backslash R}{%
  (P(x) \And Q) \leadsto R}
\end{equation}
\bop
\begin{PreuveL}{ }
1 & (\exists x\cdot (P(x))\And Q)\leadsto R &  premise\\
2 & (\exists y\cdot (P(y))\And Q)\leadsto R &  1\\
3 & P(x)\Imp \exists y\cdot (P(y)) &  for any $x$ \\
4 & P(x)\And Q\Imp (\exists y\cdot (P(y))\And Q) &  3\\
5 & (P(x)\And Q)\leadsto (\exists y\cdot (P(y))\And Q) &  4 and BRL\\
6 & (P(x)\And Q)\leadsto R &  TRA 5 and 2
\end{PreuveL}
\eop

\noindent Now, the equivalence between (\ref{tlip}) and LIP proof
obligation is as follows:
\begin{Preuve}
y\in r^{-1}[p\cap \cpl{q}]\And \neg \grd(\m{G'})\leadsto \grd(\m{G'})\\
\Com{\equiv}{ using (\ref{E1}) }\\
\exists x\cdot (P(x)\And \neg Q(x)\And I(x)\And J(y,x))\And \neg \grd(\m{G'})\leadsto \grd(\m{G'})\\
\Com{\equiv}{ $x\backslash \grd(G')$, (\ref{ThLTO}) and DSJ }\\
P(x)\And \neg Q(x)\And I(x)\And J(y,x)\And \neg \grd(\m{G'})\leadsto \grd(\m{G'})
\end{Preuve}
\eop

\section{Conclusions}\label{CONC}

In this report we present a formal model of fair iteration of events
in a \B\ event system. Moreover we use the model to justify our proof
obligations for basic liveness properties and preservation under
refinement of general liveness properties. The model of fair iteration
of events uses the dovetail operator, an operator proposed in
\cite{BrNeAFCD} to model fair choice. Our proofs are mainly founded in
fixpoint calculations of fair iteration of events and weakest
precondition calculus. 

Our approach to justify our proof obligations was inspired by
\cite{AbrMusIDCB}. The approach, founded in fixpoint calculations and
weakest precondition calculus, to justify proof obligations about
liveness properties is not classical. It is common to justify proof
obligations of this kind of properties by operational reasoning
about state traces in the system \cite{BaQiRFAS}, and the
justifications are not so formal as expected. The approach taken in this
report allows us to make axiomatic proofs and verify it with the
prover of atelier \B.

As a future work, we investigate the relationship between general
liveness properties and the iteration of events under weak fairness or
minimal progress assumptions. We are mainly interested in sufficient
conditions to guarantee preservation of liveness properties when a
system with weak fairness assumptions is refined in a system with
minimal progress assumptions.

\bibliographystyle{plain} 
\bibliography{./docs}

\end{document}